# Topological Atomic Displacements, Kirchhoff and Wiener Indices of Molecules

Ernesto Estrada[1]

Department of Mathematics, Department of Physics and Institute of Complex Systems, University of Strathclyde, Glasgow G11XQ

and

Naomichi Hatano

Institute of Industrial Science, University of Tokyo, Komaba, Meguro 153-8505, Japan

**Abstract**

We provide a physical interpretation of the Kirchhoff index of any molecules as well as of the Wiener index of acyclic ones. For the purpose, we use a local vertex invariant that is obtained from first principles and describes the atomic displacements due to small vibrations/oscillations of atoms from their equilibrium positions. In addition, we show that the topological atomic displacements correlate with the temperature factors (B-factors) of atoms obtained by X-ray crystallography for both organic molecules and biological macromolecules.

[1] Corresponding author. E-mail: ernesto.estrada@strath.ac.uk

## 1. Introduction

Many topological ideas have been introduced in chemistry in an *ad hoc* way [1]. A classical example is provided by the oldest topological index, which is nowadays known as the Wiener index, $W$ [2]. It is defined as the sum of all shortest-path distances between (non-hydrogen) atoms in a molecule. This index correlates very well with many physico-chemical properties of organic molecules [3]. Several attempts to provide a physico-chemical interpretation of $W$ have been conducted. In one of them, $W$ has been shown to represent a rough measure of the molecular surface area [4]. More recently, Gutman and Zenkevich [5] have shown that this index is related to the internal energy of organic molecules, with a special role played by the vibrational energy.

In general, very few approaches to defining topological invariants in chemistry start from first-principle physical concepts, deriving indices which are physically sound and chemically useful. An attempt to define a topological index along the line of this strategy was developed by Klein and Randić, who defined the so-called Kirchhoff index, $Kf$ [6]. The Kirchhoff index is defined in an analogous way to the Wiener index but by using the concept of resistance distance $r_{ij}$ between pairs of nodes instead of the shortest-path distance. Despite that the index uses well-known concepts from physics such as Ohm's and Kirchhoff's laws [6], it is not straightforward to realise what the "electrical resistance" means for a chemical bond. These difficulties have urged us to search for a first-principle approach to defining topological invariants with a clear physico-chemical interpretation and that solve existing chemico-structural problems.

Here we derive a local vertex invariant from first principles which describes the atomic displacements due to small vibrations/oscillations of atoms from their equilibrium positions. Using this approach we provide a clear and unambiguous physical interpretation of the Kirchhoff index of any molecule in terms of atomic displacements. We show here that the Kirchhoff index is the sum of the squared atomic displacements produced by small molecular vibrations or oscillations of atoms from their equilibrium positions. For acyclic molecules as the ones studied by Gutman and Zenkevich [5], our results explain the relationship between the Wiener index and vibrational molecular energy. The topological atomic displacements are shown here to correlate with the temperature factors (B-factors) of atoms obtained by X-ray crystallography. We illustrate our results for both organic molecules and proteins.

## 2. Background

Here we represent molecules as graphs $G=(V,E)$, where nodes represent united atoms and edges represent physical interactions between such united atoms. In the simplest case of an alkane molecule the nodes represent the united atoms $CH_n$, where $n=0,1,2,3$, and the edges are the covalent C-C bonds; in other words, the graph corresponds to the hydrogen-depleted molecular graph. However, we are not constrained here to such representation. For instance, a protein can be represented through its residue interaction graph/network [7]. In this approach the nodes are united-atom representations of the amino acids, centred at their $C_\beta$ atoms, with the exception of glycine for which $C_\alpha$ is used. Two nodes are then connected if the distance $r_{ij}$ between both $C_\beta$ atoms of the residues $i$ and $j$ is not longer than a certain cutoff value $r_C$. The elements of the adjacency matrix of the residue network are obtained by

$$A_{ij}=\begin{cases} H(r_C-r_{ij}) & i\neq j \\ 0 & i=j \end{cases},$$

where $H(x>0)=1$ and $H(x\leq 0)=0$. We use $r_C=7.0$ Å [7] hereafter.

The Wiener index $W$ is defined as [2]

$$W=\sum_{i<j} d_{ij}\,, \qquad (1)$$

where $d_{ij}$ is the shortest-path distance between atoms $i$ and $j$ in the molecular graph. In the case of a molecular network like a residue network the Wiener index divided by the number of nodes has been used as a criterion for defining 'small-world' networks.

The Kirchhoff index is defined as [6]

$$Kf = \sum_{i<j} r_{ij} , \tag{2}$$

where the resistance distance $r_{ij}$ between nodes $i$ and $j$ in a graph is obtained through the Moore-Penrose generalised inverse of the Laplacian matrix $\mathbf{L}^+$ [8], as

$$r_{ij} = \left(\mathbf{L}^+\right)_{ii} + \left(\mathbf{L}^+\right)_{jj} - 2\left(\mathbf{L}^+\right)_{ij} . \tag{3}$$

The Laplacian matrix is defined as $\mathbf{L} = \mathbf{D} - \mathbf{A}$, where $\mathbf{D}$ is the diagonal matrix of node degrees and $\mathbf{A}$ is the adjacency matrix. It is well-known that for acyclic molecules the Wiener and the Kirchhoff indices coincide.

## 3. Topological Atomic Displacements

We now consider the classical analogy in which the atoms are represented by balls and bonds are identified with springs with a common spring constant $k$ [9]. We would like to consider a vibrational excitation energy from the static position of the molecule. Let $x_i$ denote the displacement of an atom $i$ from its static position. Then the vibrational potential energy of the molecule can be expressed as

$$V(\vec{x}) = \frac{k}{2} \vec{x}^T \mathbf{L} \vec{x} , \tag{4}$$

where $\vec{x}$ is the vector whose $i$th entry $x_i$ is the displacement of the atom $i$.

Now we are going to suppose that the molecule is immersed into a thermal bath of inverse temperature $\beta = \frac{1}{k_B T}$, where $k_B$ is the Boltzmann constant. Then the probability distribution of the displacement of the nodes is given by the Boltzmann distribution

$$P(\vec{x}) = \frac{e^{-\beta V(\vec{x})}}{Z} = \frac{1}{Z}\exp\left(-\frac{\beta k}{2}\vec{x}^T \mathbf{L}\vec{x}\right), \quad (5)$$

where the normalization factor $Z$ is the partition function of the molecule

$$Z \equiv \int d\vec{x} \exp\left(-\frac{\beta k}{2}\vec{x}^T \mathbf{L}\vec{x}\right). \quad (6)$$

The mean displacement of an atom $i$ can be expressed by

$$\Delta x_i \equiv \sqrt{\left\langle x_i^2 \right\rangle} = \sqrt{\int x_i^2 P(\vec{x}) d\vec{x}} .$$

(7)

We can calculate this quantity once we can diagonalise the Laplacian matrix $\mathbf{L}$. Let us denote by $U$ the matrix whose columns are the orthonormal eigenvectors $\vec{\psi}_\mu$ and $\Lambda$ the diagonal matrix of eigenvalues $\lambda_\mu$ of the Laplacian matrix. Note here that the eigenvalues of the Laplacian of a molecular graph are positive except for one zero eigenvalue. Then, we write the Laplacian spectrum as $0 = \lambda_1 < \lambda_2 \leq \cdots \leq \lambda_n$. An important observation here is that the *zero eigenvalue does not contribute to the vibrational energy*. This is because the mode $\mu = 1$ is the mode where all the atoms (balls) move coherently in the same direction and thereby the whole molecule moves in one direction. In other words, this is the motion of the centre of mass, not a vibration.

In calculating Eqs. (5) and (6), the integration measure is transformed as

$$d\vec{x} \equiv \prod_{i=1}^{n} dx_i = \left|\det U\right| \prod_{i=1} dy_i = d\vec{y} \quad (8)$$

because the determinant of the orthogonal matrix, $\det U$, is either $\pm 1$. Then we have

$$
\begin{aligned}
Z &= \int d\vec{y} \exp\left( -\frac{\beta k}{2} \vec{y}^T \Lambda \vec{y} \right) \\
&= \prod_{\mu=1}^{n} \int_{-\infty}^{+\infty} dy_\mu \exp\left( -\frac{\beta k}{2} \lambda_\mu {y_\mu}^2 \right).
\end{aligned} \tag{9}
$$

Note again that because $\lambda_1 = 0$ the contribution from this eigenvalue obviously diverges. This is because nothing stops the whole molecule from moving coherently in one direction. When we are interested in the vibrational excitation energy within the network, we should offset the motion of the centre of mass and focus on the relative motion of the nodes. We therefore redefine the partition function by removing the first component $\mu = 1$ from the last product. We thereby have

$$
\begin{aligned}
\widetilde{Z} &= \prod_{\mu=2}^{n} \int_{-\infty}^{+\infty} dy_\mu \exp\left( -\frac{\beta k}{2} \lambda_\mu {y_\mu}^2 \right) \\
&= \prod_{\mu=2}^{n} \sqrt{\frac{2\pi}{\beta k \lambda_\mu}}.
\end{aligned} \tag{10}
$$

Next we calculate the mean displacement $\Delta x_i$ defined by Eq. (7). We first compute the numerator of the right-hand side of Eq. (7) as follows:

$$
\begin{aligned}
I_i &= \int d\vec{x} x_i^2 \exp\left( -\frac{\beta k}{2} \vec{x}^T \mathbf{L} \vec{x} \right) \\
&= \int d\vec{y} \left( U\vec{y} \right)_i^2 \exp\left( -\frac{\beta k}{2} \vec{y}^T \Lambda \vec{y} \right) \\
&= \int d\vec{y} \left( \sum_{\nu=1}^{n} U_{i\nu} y_\nu \right)^2 \exp\left( -\frac{\beta k}{2} \vec{y}^T \Lambda \vec{y} \right) \\
&= \int d\vec{y} \left( \sum_{\nu=1}^{n} \sum_{\gamma=1}^{n} U_{i\nu} U_{i\gamma} y_\nu y_\gamma \right) \prod_{\mu=1}^{n} \exp\left( -\frac{\beta k}{2} \lambda_\mu y_\mu^2 \right),
\end{aligned} \tag{11}
$$

On the right-hand side, any terms with $\nu \neq \gamma$ will vanish after integration because the integrand is an odd function with respect to $y_\nu$ and $y_\gamma$. The only possibility of a finite result is due to terms with $\nu = \gamma$. We therefore have

$$I_i = \int d\vec{y}\left[\sum_{\nu=1}^{n}(U_{i\nu}y_\nu)^2\right]\prod_{\mu=1}^{n}\exp\left(-\frac{\beta k}{2}\lambda_\mu y_\mu^2\right)$$

$$= \int_{-\infty}^{+\infty} dy_1 (U_{i1}y_1)^2 \times \prod_{\mu=2}^{n}\int_{-\infty}^{+\infty} dy_\mu \exp\left(-\frac{\beta k}{2}\lambda_\mu y_\mu^2\right) + \tag{12}$$

$$\sum_{\nu=2}^{n}\int_{-\infty}^{+\infty} dy_\nu (U_{i\nu}y_\nu)^2 \exp\left(-\frac{\beta k}{2}\lambda_\nu y_\nu^2\right) \times \prod_{\substack{\mu=1\\ \mu\neq\nu}}^{n}\int_{-\infty}^{+\infty} dy_\mu \exp\left(-\frac{\beta k}{2}\lambda_\mu y_\mu^2\right),$$

where we separated the contribution from the zero eigenvalue and those from the other ones. Due to the divergence introduced by the zero eigenvalue we proceed the calculation by redefining the quantity $I_i$ with the zero mode removed:

$$\widetilde{I}_i \equiv \sum_{\nu=2}^{n}\int_{-\infty}^{+\infty} dy_\nu (U_{i\nu}y_\nu)^2 \exp\left(-\frac{\beta k}{2}\lambda_\nu y_\nu^{\,2}\right)\times\prod_{\substack{\mu=2\\ \mu\neq\nu}}^{n}\int_{-\infty}^{+\infty} dy_\mu \exp\left(-\frac{\beta k}{2}\lambda_\mu y_\mu^{\,2}\right)$$

$$= \sum_{\nu=2}^{n}\frac{U_{i\nu}^{\;2}}{2}\sqrt{\frac{8\pi}{(\beta k\lambda_\nu)^3}}\times\prod_{\substack{\mu=2\\ \mu\neq\nu}}^{n}\sqrt{\frac{2\pi}{\beta k\lambda_\nu}} \tag{13}$$

$$= \widetilde{Z}\times\sum_{\nu=2}^{n}\frac{U_{i\nu}^{\;2}}{\beta k\lambda_\nu}.$$

We therefore arrive at the following expression for the mean displacement of an atom:

$$\Delta x_i \equiv \sqrt{\left\langle x_i^{\;2}\right\rangle} = \sqrt{\frac{\widetilde{I}_i}{\widetilde{Z}}} = \sqrt{\sum_{\nu=2}^{n}\frac{U_{i\nu}^{\;2}}{\beta k\lambda_\nu}}. \tag{14}$$

If we designate by $\mathbf{L}^+$ the Moore-Penrose generalised inverse of the graph Laplacian [8], which has been proved to exist for any molecular graph, then it is straightforward to realise that

$$(\Delta x_i)^2 = \frac{1}{\beta k}\left(\mathbf{L}^+\right)_{ii}. \tag{15}$$

## 4. Kirchhoff and Wiener indices revisited

From now on we are going to consider the case $\beta k \equiv 1$ for the sake of simplicity. Then, it is easy to see that due to the orthonormality of the eigenvectors of the inverse Laplacian, we have

$$\sum_{i=1}^{n}(\Delta x_i)^2 = tr(\mathbf{L}^+) = \sum_{j=2}^{n}\frac{1}{\lambda_j} = \frac{1}{n}Kf(G), \tag{16}$$

That is, the Kirchhoff index of a molecular graph is simply the sum of the squared atomic displacements produced by small molecular vibrations multiplied by the number of atoms in the molecule. Since it is well-known that for acyclic molecules, i.e., molecular trees, the Kirchhoff and Wiener indices coincide, we also have

$$W(T) = n\sum_{i=1}^{n}(\Delta x_i)^2 \; . \tag{17}$$

Then the potential energy (4) can be expressed as

$$V(\vec{x}) = \frac{1}{2}\sum_{i=1}^{n}k_i(\Delta x_i)^2 - \sum_{i,j\in E}(\Delta x_i)(\Delta x_j).$$

Let $R_i = \sum_{j=1}^{n} r_{ij}$ be the sum of all resistance distances from atom $i$ to any atom in the molecule, i.e., the sum of the $i$ th row (or column) of the resistance distance matrix. That is, $R_i = \sum_{j=1}^{n}\left(L_{ii}^{+} + L_{jj}^{+} - 2L_{ij}^{+}\right)$. It is known that $\sum_{j=1}^{n} L_{ij}^{+} = 0$ . Then,

$$R_i = nL_{ii}^{+} + tr(L^+) = n(\Delta x_i)^2 + \sum_{i=1}^{n}(\Delta x_i)^2 \; , \tag{18}$$

This relation indicates that $(\Delta x_i)^2$ and $R_i$ are linearly related for the atoms of a given molecule. Using Eq. (18) we can express the potential energy (4) in terms of the resistance distance of the atoms in the molecular graph

$$V(\vec{x}) = \frac{1}{2n}\sum_{i=1}^{n}k_i R_i - \frac{1}{n^2}\sum_{i,j\in E}\left[R_i R_j - \frac{Kf}{n}\left(R_i + R_j\right) + \frac{1}{n^2}Kf\right]^{1/2} - \frac{Kf}{2n}\langle k\rangle \; . \tag{19}$$

where $\langle k\rangle$ is the average degree of the molecular graph. The first term in the right-hand side of Eq. (19) was already introduced by Estrada et al. [10] as a topological index obtained from the quadratic form $\langle \mathbf{v} | \mathbf{D} | \mathbf{u} \rangle$, where $\mathbf{v}$ is a vector of node degrees, $\mathbf{D}$ is the distance matrix and $\mathbf{u}$ is an all-one vector.

In summary, the normalised Kirchhoff index of a molecular graph represents the sum of squared displacements of atoms due to molecular vibrations and the sum of resistance distances for a given atom depends linearly on the square of the displacement of the corresponding atom. The term $(\Delta x_i)^2$ has a very clear physical interpretation. It represents the atomic displacement due to molecular vibrations. Small values of $(\Delta x_i)^2$ indicate that those atoms are very rigid in the molecule. For instance, in 2,2,3-trimethylbutane the smallest displacement is obtained for the carbon atom connected to three methyl groups $\Delta x_C = 0.534$, followed by the one bounded to two $CH_3$ groups, $\Delta x_{CH} = 0.655$. Then, the methyl groups display the largest displacements, $\Delta x_{CH_3} = 1.000$ for those at position 2 and $\Delta x_{CH_3} = 1.069$ for those at position 3.

## 5. Topological Displacements and Temperature Factors

We guess that the atomic displacement $\Delta x_i$ should display some linear correlation with an experimental measure of how much an atom oscillates or vibrates around its equilibrium position. Such experimental measure is provided by X-ray experiments as the so-called B-factor or temperature factor, and represents the reduction of coherent scattering of X-rays due to the thermal motion of the atoms. For instance, in the molecule of naphthalene the atomic displacements of carbon atoms correlate very well with the experimental B-factors (in parenthesis) [11]: 0.898 (4.6 Å$^2$), 0.815 (4.0 Å$^2$) and 0.615 (3.4 Å$^2$), which gives a correlation coefficient $r = 0.96$. The following correlation coefficients are obtained for: anthracene [12] ($r = 0.99$); phenanthrene [13] ($r = 0.95$); pyrene [14] ($r = 0.99$); and triphenylene [15] ($r = 0.99$). In these cases we averaged the values of B-factors for equivalent carbon atoms. In Fig. 1 we plot the values of $\Delta x_i$ versus the B-factors for the carbon atoms of anthracene and pyrene.

**Insert Fig. 1 about here.**

The B-factors are quite relevant for the study of protein structures as they contain valuable information about the dynamical behaviour of proteins and several methods have been designed for their prediction [16]. It is known that regions with large B-factors are usually more flexible and functionally important. The atomic displacements have been used previously by Bahar et al. [17] to describe thermal fluctuations in proteins. We note in passing that we use here a residue network representation of the protein based on $\beta$-carbons instead of the $\alpha$-carbons used by Bahar et al.

For the sake of illustration we have selected here the lipase b from *Candida antarctica* (1tca) [18]. In this case we obtain a correlation coefficient $r = 0.74$ between the experimental B-factors and the topological atomic displacements. For this protein Yuan et al. [19] reported $r = 0.63$ for predicting the experimental B-factors. In Fig. 2 (top) we illustrate the profiles for the normalised B-factors and the topological atomic displacements of residues for this protein. We also represent in Fig. 2 (bottom) the 20 residues with the highest values of $\Delta x_i$ in the molecular structure of the protein. We recall that the residues with the largest values of the atomic displacements are those displaying the highest flexibility in the protein. Here we have represented these residues by using blue colour for the atoms in these residues. We also represent the 20 residues with the lowest values of $\Delta x_i$, which correspond to those displaying the highest rigidity in the protein. They are coloured in red in the molecular structure of the protein. As can be seen the most flexible amino acids are those which are on the surface of the protein, while the most rigid ones are concentrated around the protein core.

**Insert Fig. 2 about here.**

The new relationship obtained here between the topological atomic displacements and the sum of the resistance distances for a given atom, i.e., the expression (17), opens up new possibilities for interpreting $\Delta x_i$ in a given molecule. According to Eq. (17) the topological displacements for the atoms in a molecule depend only on the sum of the resistance distances for the corresponding atom, e.g., $(\Delta x_i)^2 \sim \frac{1}{n}\sum_j r_{ij}$ [6]. It is known that if there is more than one path connecting two atoms in a molecule, i.e., there are cycles, the resistance distance is smaller than in the case when there is only a single path. Then, if there is one oscillation/vibration in one atom which is transmitted to all the other atoms through the different paths connecting them, the vibration is attenuated along every path. Consequently, a small value of $\Delta x_i$ is due to the fact that the atom $i$ is part of a large number of paths connecting it to other atoms. This implies that when the other atoms oscillate/vibrate their effect is very much attenuated before arriving to $i$.

## 6. Conclusions

We have developed a theoretical approach based on classical molecular mechanics to accounting for small displacements of atoms from their equilibrium positions due to oscillations or vibrations. The topological atomic displacements are expressed in terms of the eigenvalues and eigenvectors of the discrete Laplacian matrix of the molecular graph. Using this approach we have given a clear and unambiguous physical interpretation of the Kirchhoff index as well as of the Wiener index of acyclic molecules. It explains previous empirical results clearly, showing that these indices are related to vibrational energy of alkanes and dithioderivative compounds. More importantly, the topological atomic displacements are well correlated with the B-factors obtained by X-ray crystallography.

**Acknowledgements**

EE thanks partial financial support from the New Professor's Fund given by the Principal, University of Strathclyde.

## Figure Captions

**Fig. 1**. Linear correlation between the topological atomic displacements and experimental B-factors for carbon atoms of anthracene (empty circles) and pyrene (empty squares). The temperature factors of equivalent atoms were averaged.

**Fig. 2**. Profiles of the topological atomic displacements (solid line) and the B-factors (dotted line) for lipase b from *Candida antarctica*, 1tca (top), and illustration of the amino acids having the 20 largest (blue) and 20 smallest (red) values of the topological atomic displacements (bottom).

**Fig. 1**

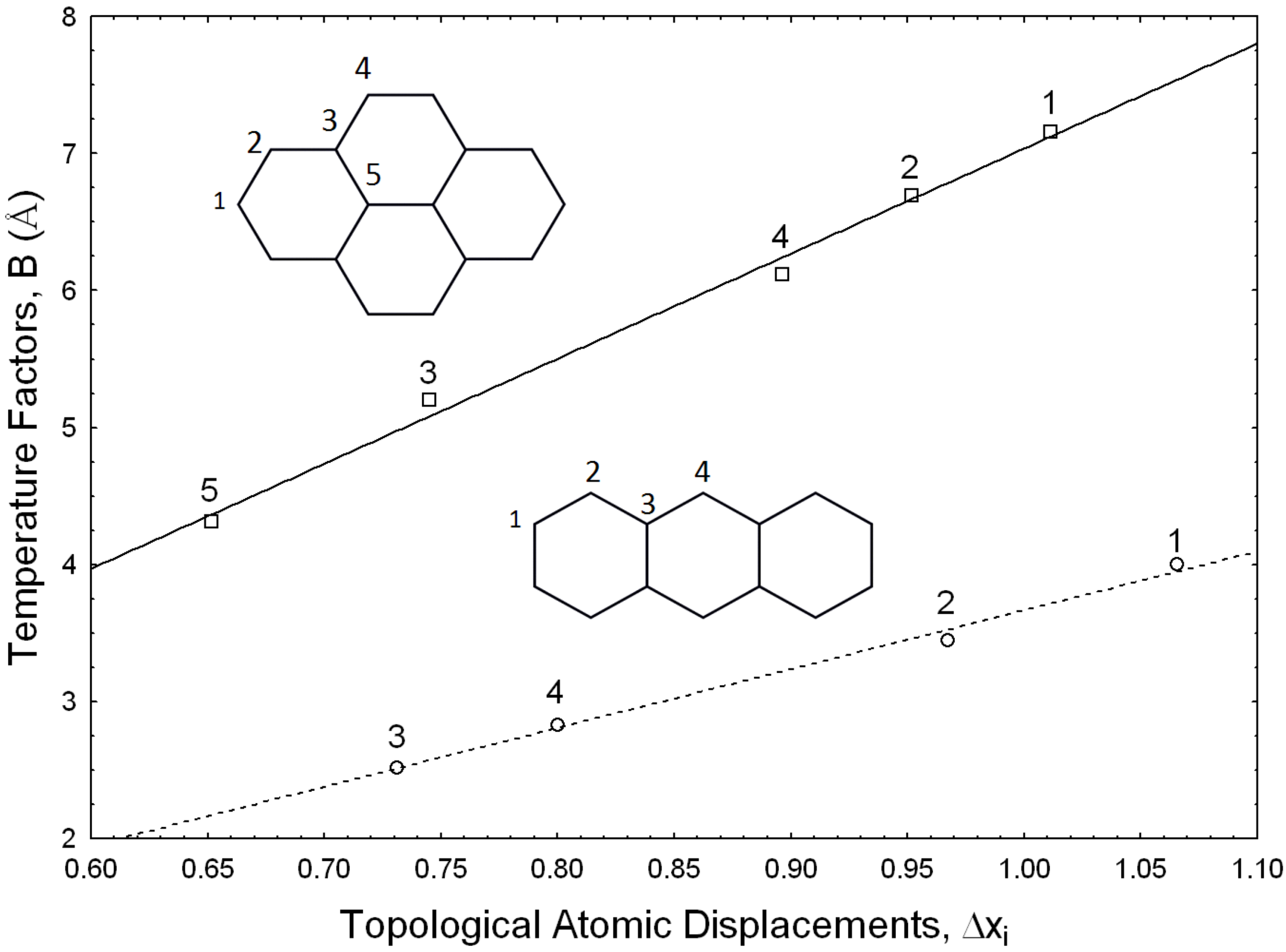
Temperature Factors, B (Å)
Topological Atomic Displacements, Δxi
8
7
6
5
4
3
2
0.60
0.65
0.70
0.75
0.80
0.85
0.90
0.95
1.00
1.05
1.10
1
2
3
4
5

**Fig. 2**

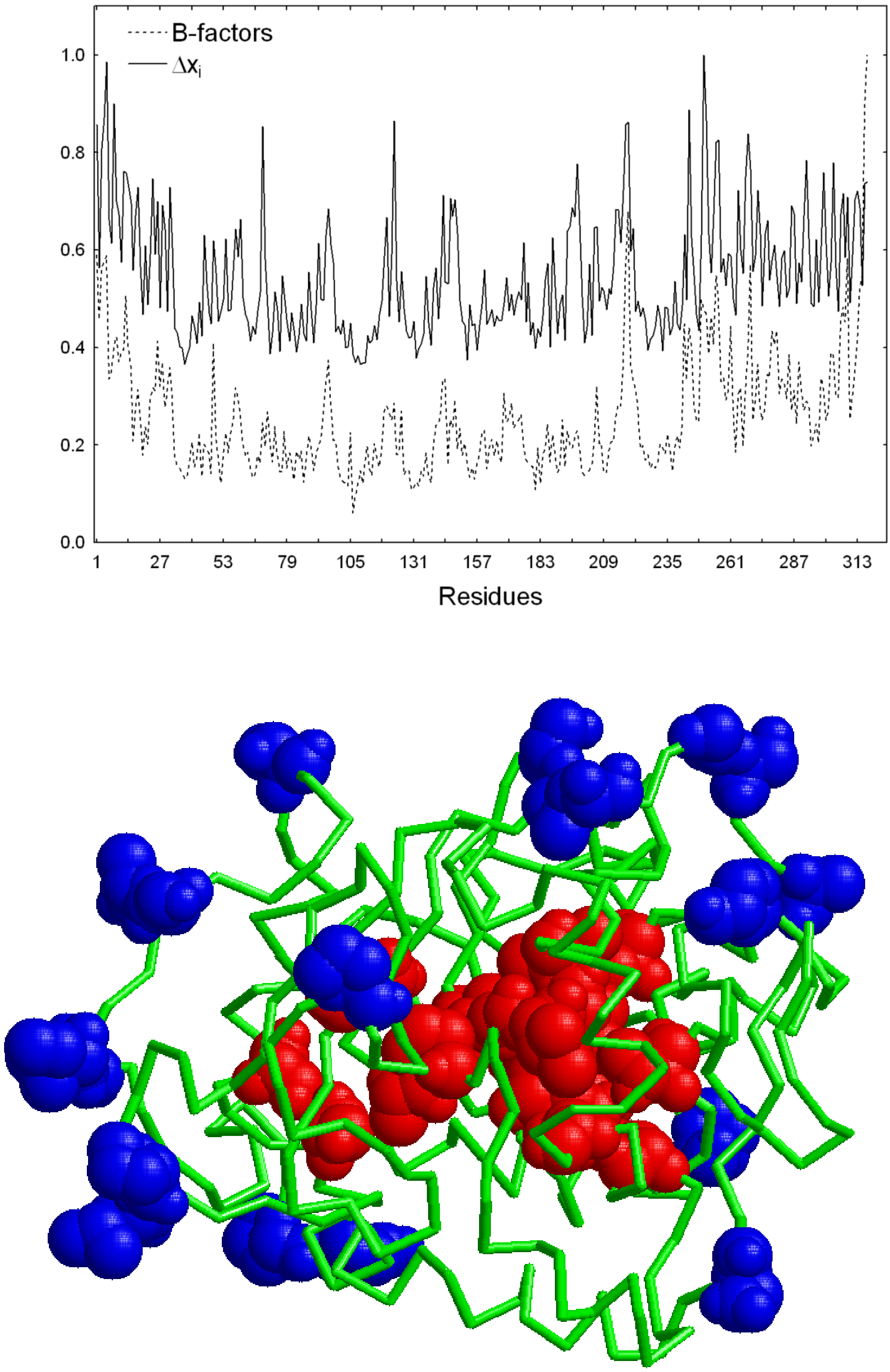

B-factors
Δx_i
1.0
0.8
0.6
0.4
0.2
0.0
1
27
53
79
105
131
157
183
209
235
261
287
313
Residues